\documentstyle[12pt,epsfig,titlepage]{article}
\begin{document}

\begin{titlepage}

\vspace*{6cm}
\begin{center}
{\LARGE\bf Calculation of the light-shifts in the} \\
{\LARGE\bf $ns$-states of hydrogenic systems} \\
\vspace{7ex}
{\large Viktor Yakhontov\dag}, {\large  Klaus Jungmann} and
{\large ************}  \\[2.5ex]
Physikalisches Institut der Universit\"{a}t Heidelberg\\
Philosophenweg 12, D-69120 Heidelberg \\
Germany
\\
\vspace{7ex}
PASC numbers:  32.60;32.80;36.10
\end{center}
\vspace{7cm}
\dag Permanent address: St.Petersburg State Technical University,
Polytechnicheskaya 29, 195251, St. Petersburg, Russia
\end{titlepage}
\section{Introduction}
\label{sec:int}

Calculation of the light-shifts of both the ground and excited states
of the atoms are mostly easily performed in the formalism of the
so-called ``dressed'' atom (see e.g.  \cite{CTandD1972}).  Within the
framework of this approach the shift, $\Delta \varepsilon_n$, of the
level $n$ occurring due to the incident photon beam of the energy,
$\omega$, and polarization, $\mbox{\boldmath $e$}$, is defined as
follows:
\begin{equation}
  \label{glshift}
  \Delta \varepsilon_n(\omega)= \frac{P}{2\varepsilon_0 \hbar^2 S c} \Re
  \sum_{r} \hspace{-14.5pt} \int \left \{ \frac{\langle n|
    \mbox{\boldmath $D \cdot e^\ast$} |r \rangle \langle
    r|\mbox{\boldmath $D \cdot e$}|n \rangle } {\varepsilon_n +\omega
    - \varepsilon_r} + \frac{ \langle n|\mbox{\boldmath $D \cdot e$}|r
    \rangle \langle r|\mbox{\boldmath $D \cdot e^\ast$}|n \rangle }
  {\varepsilon_n - \omega - \varepsilon_r } \right\}.
\end{equation}
Here $\mbox{\boldmath $D$}=e \mbox{\boldmath $r$}$ is the dipole
momentum operator of the electron; $P=\frac{1}{2} \varepsilon_0 S c E^2$ is
the power of the photon beam with the cross section $S$ and the
electric strength $E$; $\Re$ designates the real part. According
to (\ref{glshift}), the calculation of the shift,
$\Delta \varepsilon_n$, reduces to that of the tensor of dynamical
polarizability (DP), $\alpha_n^{ij}(\omega)$,
of the state $n$.  It is defined as (see e.g. \cite{LP})
\begin{equation}
  \label{alpha}
  \alpha_n^{ij}(\omega)=- \sum_{r} \hspace{-14.5pt} \int \left \{
  \frac{\langle n|D_i|r \rangle \langle r|D_j |n \rangle }
  {\varepsilon_n +\omega - \varepsilon_r} + \frac{ \langle n|D_j|r
    \rangle \langle r|D_i |n \rangle } {\varepsilon_n - \omega -
    \varepsilon_r } \right\}.
\end{equation}
Here $D_i$ denotes the $i$th component of the vector of the dipole
moment. An {\em ab initio\/} exact analytical calculation of DP is
evidently possible for hydrogenic systems only and the corresponding
results are known (in principle) for all states already for long time
(see \cite{RZM} and references herein).  They are usually performed by means of
the exact explicit expression
for the non-relativistic Green's function either in coordinate or
momentum spaces. As a result, DP are expressed in terms of
the special (Appel) functions whose complicated mathematical structure
makes accurate numerical (and analytical) analysis of
$\alpha_n^{ij}(\omega)$ hard to carry out. Especially this argument
refers to the case when the photon energy, $\omega$, lies in the
vicinity of the threshold: $\omega \sim I_n,\;\:I_n$ being the
ionization potential of the state $n$.  It is not surprising,
therefore, that calculations of this type seem to be available for the
hydrogenic $1s$-state only (see \cite{Gav}). In fact, provided that
the principal quantum number, $n$, of the level is fixed,
$\alpha_n^{ij}(\omega)$ has singularities when the photon energy,
$\omega$, is in the resonance with the higher/lower discrete levels of
the atom \footnote{This singularities can be avoided if the finite
widths of all atomic levels are taken into account.}:
$\omega=|\varepsilon_i-\varepsilon_n|,\;i>n, i<n$.  Beyond that, DP
acquires also a non-zero imaginary part if $\omega \geq I_n$.  For
such $\omega$ the real part of DP, $\Re\alpha_n^{ij}(\omega)$,
describes (as for $\omega < I_n$) the shift of the level, whereas
imaginary part, $\Im\alpha_n^{ij}(\omega)$, allows for a decay
probability (photoionization) of the atom under the action of the
photon field. Namely, according to the optical theorem \cite{LP}:
$\sigma_n^{(\gamma)}(\omega)=4\pi\alpha\omega
\Im\alpha_n^{ij}(\omega)$, where $\sigma_n^{(\gamma)}\omega)$ is the
total photoionization cross section of the state with the principal
quantum number, $n$, and $\alpha=e^2/(\hbar c)=1/137$ is the fine
structure constant.  In mathematical terms, $\alpha_n^{ij}(\omega)$
has unremovable singularity at the point $\omega=I_n$, so that a great
care must be taken to make numerical calculation within this region of
$\omega$ stable and highly accurate. The most efficient way of
achieving that consists in combining numerical methods together with
analytical ones. It is relevant to point out that the detailed
description of the DP's behavior of the mentioned type proves to be of
particular importance, e.g. for the problem of the light-shifts'
calculation in muonium atom, denoted $\left(\mu^+-e^-\right)^0$. This
is due to the fact that in the highly accurate experimental
measurements of the $1s-2s$ energy splitting in this exotic system,
which are in progress now, the energies,  $\omega_1$ and $\omega_2$, of two
incident photon beams are supposed to be in the resonance with the
following transitions: $1s + \hbar \omega_1 + \hbar
\omega_1 \rightarrow 2s$ and $2s + \hbar \omega_2 \rightarrow
\varepsilon p$ (see \cite{Jung} for more details).  Therefore precise
calculation of the corresponding light-shifts of $1s$- and $2s$-levels
would be rather desirable.

In the current letter we present the results of analytical and
numerical calculation of $\alpha_n^{ij}(\omega)$ together with the
corresponding light-shifts in the $ns$-states, $n=1,2$, of the
muonium atom. It should be pointed out that the similar calculation by
Beausoleil \cite{Beausoleil} employing a pure numerical scheme proves
to be incomplete. Besides, in the contrast to
the usual technique (i.e.  by means of the Green's function) the
current calculation is carried out in the fashion of Sternheimer
\cite{Sternheimer} where (exact and analytical) summation over the
intermediate states $r$ in (\ref{alpha}) is reduced to solution of a
certain differential equation. Such
an approach, which is applied to the problem under consideration for
the first time, to our knowledge, seems to be rather instructive.
Apart from its self-contained academic interest, it may also give certain
advantages in treating the higher $ns$-states ($n \geq 4$) of the hydrogenic
systems, as well as for exact calculation of the various
$\omega$-dependent sums of the form:
\begin{equation}
  \label{gensum}
  S_n^{(\mu)}(\omega)= \sum_{s} \hspace{-14.5pt} \int \left \{
  \frac{\langle n||r ||s \rangle \langle s||r^{\mu} ||n \rangle
    } {\varepsilon_n +\omega - \varepsilon_s} + \frac{ \langle
    n||r||s \rangle \langle s||r^{\mu}||n \rangle }
  {\varepsilon_n - \omega - \varepsilon_s } \right\}.
\end{equation}
Here $\mu$ is an arbitrary number, being not necessarily
positive and integer; $\langle s||r^{\mu}||n \rangle$ denotes reduced
matrix element. Such type of expressions make their appearance in
numerous problems of atomic physics.

\section{Calculation of the light-shifts}
\subsection{General consideration}

For the $ns$-states under consideration Eq.(\ref{glshift}) can
be reduced to the following angular- and spin-independent form
\cite{Sandars}
\begin{equation}
  \label{sclshift}
  \Delta \varepsilon_{ns}(\omega)= -\frac{P}{2\varepsilon_0 \hbar^2 S c}
  (\mbox{\boldmath $e \cdot e^\ast$}) \Re \alpha_{ns}^{S}(\omega) \equiv
  -\frac{P}{2\varepsilon_0 \hbar^2 S c}\Re \alpha_{ns}^{S}(\omega).
\end{equation}
Here $\alpha_{ns}^{S}(\omega)$ denotes the so-called scalar DP
(henceforth the atomic units, $e^2=\hbar=m_e=1$ are used):
\begin{equation}
  \label{scpol}
  \alpha_{ns}^{S}(\omega)= -\frac{1}{3} \sum_{kp} \hspace{-14.5pt}
  \int \left \{ \frac{\langle ns|| r ||kp \rangle \langle kp|| r ||ns
    \rangle } {\varepsilon_{ns} +\omega - \varepsilon_{kp}+ i0}
  + \frac{\langle ns|| r ||kp \rangle \langle kp|| r ||ns \rangle }
  {\varepsilon_{ns} - \omega - \varepsilon_{kp} } \right\},
\end{equation}
which involves the radial integrals only. Summation is performed here
over complete set of discrete and continuum $p$-states of the Coulomb
field with the charge $Z$. The infinitesimal positive imaginary
constant added in the denominator of the first term in the sum defines
the sign of $\Im\alpha_{ns}^{S}(\omega)$ occurring if
$\omega>|\varepsilon_{ns}|$, $|\varepsilon_{ns}|=Z^2/(2n^2)$
being the ionization potential of the $ns$-state. Calculation of
$\alpha_{ns}^{S}(\omega)$, Eq.(\ref{scpol}), is actually the final aim of
our consideration.

Let us introduce auxiliary function $\psi_n(r;E)$ by the equation:
\begin{equation}
  \label{psi}
  \psi_n(r;E)=\sum_{kp} \hspace{-14.5pt} \int \frac{|kp\rangle \langle
    kp|| r ||ns \rangle } {E-\varepsilon_{kp}},
\end{equation}
$E=\varepsilon_{ns} \pm \omega +i0$ being a parameter. In terms
of $\psi_n(r;E)$ DP, $\alpha_{ns}^{S}(\omega)$, is expressed as:
\begin{equation}
  \label{alpsi}
  \alpha_{ns}^{S}(\omega)=-\frac{1}{3}\left[ \langle ns|| r
  ||\psi_n(\varepsilon_{ns}+\omega+i0) \rangle + \langle ns|| r
  ||\psi_n(\varepsilon_{ns}-\omega) \rangle \right].
\end{equation}
By acting on $\psi_n(r)$ of Eq.(\ref{psi}) with the operator,
\[
  \label{oper}
  E-\widehat{H}^{(l=1)} \equiv
  E+\frac{1}{2r^2}\frac{\partial}{\partial r} \left( r^2
  \frac{\partial}{\partial r}\right) -\frac{1}{r^2} + \frac{Z}{r},
\]
and by virtue of completeness of the set $|kp\rangle$ one immediately
obtains the following inhomogeneous differential equation obeyed by
the function $\psi_n(r)$:
\begin{equation}
  \label{eqpsi}
  \frac{1}{r^2}\frac{\partial }{\partial r} \left( r^2 \frac{\partial
    \psi_n}{\partial r}\right) +2\left(E-\frac{1}{r^2} +
  \frac{Z}{r}\right) \psi_n =2 r R_{ns}(r).
\end{equation}
Here $R_{ns}(r)$ denotes the radial non-relativistic coulomb
$s$-function. In terms of the new parameters,
\begin{equation}
  \label{varch}
  \nu=\frac{Z}{\sqrt{-2E}},\;\;\; \rho=\frac{2Z}{\nu} r,
\end{equation}
Eq.(\ref{eqpsi}) takes the form:
\[
  \label{eqpsirho}
  \frac{1}{\rho^2}\frac{\partial }{\partial \rho} \left( \rho^2
  \frac{\partial \psi_n}{\partial \rho}\right)
  +\left(-\frac{1}{4}-\frac{2}{\rho^2} + \frac{\nu}{\rho}\right) \psi_n
  =\frac{1}{4}\left(\frac{\nu}{Z}\right)^3 \rho R_{ns}\left(\frac{\nu
    \rho}{2Z}\right).
\]
Finally, on introducing the new auxiliary function $\zeta(\rho)$ as
\begin{equation}
  \label{zeta}
  \psi_n(r)=\rho e^{-\rho/2} \zeta(\rho),
\end{equation}
we result in the equation of the hypergeometric type,
\begin{equation}
  \label{eqzeta}
  \rho \zeta''(\rho) + (4-\rho) \zeta'(\rho) + (\nu-2) \zeta(\rho)=
  \frac{1}{4}\left(\frac{\nu}{Z}\right)^3 \rho e^{\rho/2}
  R_{ns}\left(\frac{\nu \rho}{2Z}\right).
\end{equation}
Its solution, $\zeta(\rho)$, is supposed to be subject
for $\Re \nu>-n$ ($n$ is fixed) to conditions:
\begin{equation}
  \label{cond}
  \zeta(\rho) = O(1),\;\mbox{as}\;\rho \rightarrow 0,\;\;\;
  \zeta(\rho) = o\left[\exp{\left(\frac{n+\nu}{2n} \rho\right)}\right],
\;\mbox{as}\;\;
\rho \rightarrow \infty.
\end{equation}
These follow directly from the definitions, Eqs.(\ref{psi}),(\ref{zeta}),
since (see \cite{LL}): $R_{ns}(r) \asymp e^{-Zr/n}$,
as $r \rightarrow \infty$,
and $R_{ns}(r)=O(1),\;R_{kp}(r)=O(r)$, as $r \rightarrow 0$.
It should be emphasized that
Eqs.(\ref{cond}) are consistent with the ``orthogonality-condition'',
\begin{equation}
  \label{orthcond}
\langle 2p | \psi_1 \rangle =
\frac{\langle 2p || r || 1s \rangle} {E-\varepsilon_{2p}},\;\mbox{if}\;n=1,
\;\mbox{or}\;\;\;
  \langle np | \psi_n \rangle = \frac{\langle np || r || ns \rangle} {E-
    \varepsilon_{np}},\;\mbox{if}\; n \geq 2,
\end{equation}
which follows from the Eq.(\ref{psi}) and by virtue of the
orthogonality of the $kp$-functions:
\[
\langle kp || r || k'p \rangle = \delta_{k,k'}.
\]
Relations (\ref{orthcond}) may be used as an additional check of
correctness of the function $\psi_n(r)$.

The general solution of (\ref{eqzeta}) has the form:
\begin{equation}
  \label{gensol}
  \zeta(\rho)=D_1 \Phi(2-\nu,4; \rho) + D_2 \Psi(2-\nu,4; \rho) +
  \zeta_0(\rho).
\end{equation}
Here $\Phi(2-\nu,4; \rho),\;\Psi(2-\nu,4; \rho)$ are the regular and
irregular solutions  of  homogeneous hypergeometric
equation \cite{AS}; $D_1,D_2$ are some arbitrary constants which will
finally be chosen to comply with (\ref{cond}); $\zeta_0(\rho)$ is some
particular solution of Eq.(\ref{eqzeta}). For the distinguished case:
$\nu=2$, explicit general solution of Eq.(\ref{eqzeta}) reads:
\[
  \zeta_{\nu=2}(\rho)= -\frac{1}{6 \rho^3} D_1 \left ( 2 e^\rho + \rho
  e^\rho + \rho^2 e^\rho + \rho^3 \mbox{Ei}(-\rho) \right) + D_2 +
  \zeta_0(\rho).
\]
Here $\mbox{Ei}(-\rho)$ stands for the integral exponential function.

Let us seek $\zeta_0(\rho)$ in the form:
\begin{equation}
  \label{laplace}
  \zeta_0(\rho) = \frac{1}{2 \pi i} \oint_\gamma e^{\rho t}
  \xi(t) dt.
\end{equation}
\begin{figure}[h]
  \centering \epsfig{file=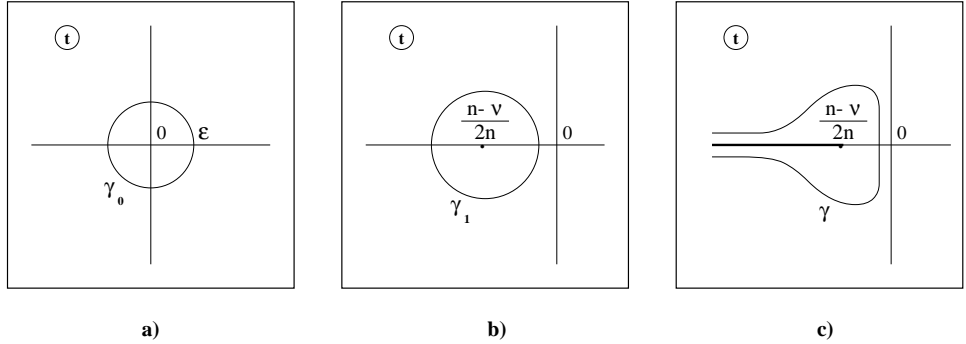}
\caption{The contours of integration in Eqs.(20),(21) and (24).}
\label{fig012}
\end{figure}
Here the integral is taken along some contour $\gamma$ in the complex
plane of $t$. It has to be chosen to comply both with (\ref{cond}) and
the form of the contour $\gamma_1$ (see below) in the integral
representation of the right-hand side of (\ref{eqzeta}).  Besides,
after passing along $\gamma$ the integrand of (\ref{laplace}) should
return back to its initial value. To find $\xi(t)$ and establish
$\gamma$ we can write first (see e.g.  \cite{LL}):
\[
  \frac{1}{4}\left(\frac{\nu}{Z}\right)^3 \rho e^{\rho/2} R_{ns}
  \left( \frac{\nu \rho}{2Z} \right) \equiv \frac{1}{2} \left(
    \frac{\nu}{\sqrt{nZ}} \right)^{3} \rho \exp{\left(\frac{n-\nu}{2n}
      \rho \right)} \Phi(-n+1,2;\nu\rho/n).
\]
Then, we use the well-known integral representation \cite{AS} of
the confluent hypergeometric function, $\Phi(-n+1,2;\nu\rho/n)$:
\begin{equation}
  \label{inhyper}
  \Phi(-n+1,2;\nu\rho/n)=-\frac{1}{2\pi i n} \oint_{\gamma_0}
  \exp{(\nu \rho t/n)} (-t)^{-n} (1-t)^n dt,
\end{equation}
where the contour $\gamma_0$, shown in the Figure \ref{fig012}(a), is
passed in the counterclockwise sense along the circle of an arbitrary
radius, $\varepsilon$. By means of Eq.(\ref{inhyper}) and after two variable
changes the right-hand side of (\ref{eqzeta}) can be finally express as
\begin{equation}
  \label{finrhside}
  \frac{1}{4}\left(\frac{\nu}{Z}\right)^3 \rho e^{\rho/2} R_{ns}
  \left( \frac{\nu \rho}{2Z} \right)= \frac{1}{4\pi i} \left(
    \frac{\nu}{\sqrt{nZ}} \right)^3 \oint_{\gamma_1} e^{\rho t}
    \left(t- \frac{n-\nu}{2n}\right)^{-n-1} \left(t-
      \frac{n+\nu}{2n}\right)^{n-1} d t.
\end{equation}
It has thereby the form similar to Eq.(\ref{laplace}).  Here the contour
$\gamma_1$, drawn in the Figure \ref{fig012}(b), is passed in the
counterclockwise sense around the point $(n-\nu)/(2n)$ along a circle
of an arbitrary radius. A substitution of (\ref{laplace}) and
(\ref{finrhside}) in Eq.(\ref{eqzeta}) yields:
\[
  \label{eqxi}
  t(1-t) \xi'(t) + (2t + \nu -1) \xi(t)=
  \frac{1}{2}\left(\frac{\nu}{\sqrt{nZ}}\right)^3 \left(t-
  \frac{n-\nu}{2n}\right)^{-n-1} \left(t-
  \frac{n+\nu}{2n}\right)^{n-1}.
\]
The general solution of this equation can be written as
\begin{equation}
  \label{solxi}
  \xi(t)=C_0 t^{1-\nu} (1-t)^{1+\nu} -
  \frac{1}{2}\left(\frac{\nu}{\sqrt{nZ}}\right)^3 t^{1-\nu}
  (1-t)^{1+\nu} \int_t^\infty t^{\nu-2} (1-t)^{-\nu-2} \frac{ \left(t-
    \frac{n+\nu}{2n}\right)^{n-1}} { \left(t-
    \frac{n-\nu}{2n}\right)^{n+1}} dt.
\end{equation}
Here $C_0$ is the arbitrary constant which can be set to $0$ for
convenience.  The integral is assumed to be taken along any path in
the complex $t$-plane which does not pass through the points:
$\{0,1,(n-\nu)/(2n)\}$.

If combined with ({\ref{laplace}), Eq.(\ref{solxi}) defines desired
particular solution, $\zeta_0(\rho)$, of Eq.(\ref{eqzeta}):
\begin{equation}
  \label{finpareq}
  \zeta_0(\rho)= - \frac{1}{2\pi i}
  \frac{1}{2}\left(\frac{\nu}{\sqrt{nZ}}\right)^3 \oint_\gamma e^{\rho
    t} t^{1-\nu} (1-t)^{1+\nu} \int_t^\infty t^{\nu-2} (1-t)^{-\nu-2}
  \frac{ \left(t- \frac{n+\nu}{2n}\right)^{n-1}} { \left(t-
    \frac{n-\nu}{2n}\right)^{n+1}} dt.
\end{equation}
In view of the given above argument, the contour $\gamma$ in
(\ref{finpareq}), is still a free ``parameter'', provided that it is
deformable into $\gamma_1$. Let us show how we can fix it.  In fact,
the integrand in Eq.(\ref{finpareq}) is analytical function in the
$t$-plane cut along any path with the ends at
$t=\infty,\;t_0=(n-\nu)/(2n)$, the latter being its logarithmic
branching point and the pole of the $n$th order at the same time. Owing
to that, we can choose $\gamma$ to be a curve which starts at
$-\infty$ at the lower edge of the cut, runs along the real axis,
encircles the point $t_0$ in the counterclockwise sense and runs back
to $-\infty$ along the upper edge of the cut (see Figure
\ref{fig012}(c)).  Such a contour is topologically equivalent to
$\gamma_1$ in the Figure \ref{fig012}(b). Moreover, after passing
along $\gamma$ the integrand returns back to its initial value, since
it decreases exponentially as $t\rightarrow -\infty$.  Here we have
temporary assumed that $\Re(\nu)> n,\;\Im\nu=0$. This restriction will
however be released later by means of the analytical continuation in
$\nu$.  For the contour $\gamma$ under consideration the integral
(\ref{finpareq}) can be split into two (independent) parts:
({\em i\/}) along two edges of
the cut and ({\em ii\/}) along the circle centered at $t_0$. The part
({\em i\/}) is reduced in its turn to the integral of the jump at the cut
of the integrand of Eq.(\ref{finpareq}). It has the form:
\begin{eqnarray}
  \lefteqn{ \int_{cut} \ldots dt = - \frac{1}{2\pi i}\frac{1}{2}
    \left(\frac{\nu}{\sqrt{nZ}}\right)^3
\int_{\frac{\nu-n}{2n}}^{\infty} e^{-\rho x} (-x)^{1-\nu}
  (1+x)^{1+\nu} \times} \nonumber \hspace{75ex} \\
\times \left \{ \left. \left.
    \int_t^\infty t^{\nu-2} (1-t)^{-\nu-2}
    \frac{\left(t- \frac{n+\nu}{2n}\right)^{n-1}} { \left(t-
      \frac{n-\nu}{2n}\right)^{n+1}} dt \right|_{t=-x-i 0} -
    (\ldots)\right|_{t=-x+i 0} \right \} dx \nonumber \\
\lefteqn{
  = -\frac{1}{2} \left(\frac{\nu}{\sqrt{nZ}}\right)^3 \frac{1}{n!}
  \frac{d^n}{dt^n} \left. \left[ t^{\nu-2} (1-t)^{-\nu-2} \left(t-
  \frac{n+\nu}{2n}\right)^{n-1} \right] \right|_{t=t_0}\times }
\nonumber \hspace{75ex} \\
\label{jump}
\times
\int_{\frac{\nu-n}{2n}}^{\infty} e^{-\rho x} (-x)^{1-\nu} (1+x)^{1+\nu} dx.
\end{eqnarray}
Conversely, the part ({\em ii\/}) is expressed as a residue of the
integrand at $t=t_0$:
\renewcommand{\arraystretch}{0.5}
\begin{eqnarray}
  \label{residue}
  -\frac{1}{2} \left(\frac{\nu}{\sqrt{nZ}}\right)^3
\begin{array}[t]{c}
  \mbox{res} \\ {\scriptstyle t=t_0}
\end{array}
\left [ e^{\rho t} t^{1-\nu} (1-t)^{1+\nu} \int_{t}^{\infty}
  t^{\nu-2} (1-t)^{-\nu-2} \frac{ \left(t-
    \frac{n+\nu}{2n}\right)^{n-1}} { \left(t-
    \frac{n-\nu}{2n}\right)^{n+1}} dt \right ].
\end{eqnarray}
Combining Eqs.(\ref{jump}) and (\ref{residue}) together we get:
\begin{eqnarray}
  \lefteqn{ \zeta_0(\rho) = -\frac{1}{2}
    \left(\frac{\nu}{\sqrt{nZ}}\right)^3 \left \{ \frac{1}{n!}
      \frac{d^n}{dt^n} \left. \left[ t^{\nu-2} (1-t)^{-\nu-2} \left(t-
      \frac{n+\nu}{2n}\right)^{n-1} \right] \right|_{t=t_0} \times \right.}
    \nonumber \hspace{75ex} \\
  \label{finzeta}
  \times \int_{\frac{\nu-n}{2n}}^{\infty} e^{-\rho x} (-x)^{1-\nu}
  (1+x)^{1+\nu} dx + \nonumber \hspace{35ex} \\ + \left.
\begin{array}[t]{c}
  \mbox{res} \\ {\scriptstyle t=t_0}
\end{array}
\left [ e^{\rho t} t^{1-\nu} (1-t)^{1+\nu} \int_{t}^{\infty} t^{\nu-2}
  (1-t)^{-\nu-2} \frac{ \left(t- \frac{n+\nu}{2n}\right)^{n-1}} {
    \left(t- \frac{n-\nu}{2n}\right)^{n+1}} dt \right ] \right\}.
\end{eqnarray}
By virtue of the integral representation \cite{AS} of the function
$\Psi(\alpha,\gamma;z)$, entering Eq.(\ref{gensol}),
\begin{equation}
  \label{PSI}
  \Psi(\alpha,\gamma;z)=\frac{1}{\Gamma(\alpha)} \int_{0}^\infty
  e^{-zt} t^{\alpha-1} (1+t)^{\gamma-\alpha-1} dt,\;\;\; \Re \alpha>0,
\end{equation}
we can finally write:
\begin{equation}
\label{finRESH}
\psi_n(r)=\rho e^{-\rho/2} \left ( D_1 \Phi(2-\nu,4; \rho) + D_2
\Psi(2-\nu,4; \rho) + \widetilde{\zeta}_0(\rho) \right).
\end{equation}
Here we have redefined the constant $D_2$ without changing its
notation; $\widetilde{\zeta}_0(\rho)$ is obtained from $\zeta_0(\rho)$
of Eq.  (\ref{finzeta}) by means of the substitution:
\begin{equation}
  \label{podst}
  \int_{\frac{\nu-n}{2n}}^{\infty} \ldots dt \rightarrow -
  \int_0^{\frac{\nu-n}{2n}} \dots dt.
\end{equation}
It it clear now that for $\nu:\:2-\nu \neq 0,-1,-2,\ldots$ the
function $\psi_n(r)$ of Eqs.(\ref{finzeta}),(\ref{podst})
will satisfy  conditions (\ref{cond}) if we set: $D_1=D_2=0$,
since \cite{AS}
\begin{eqnarray*}
  \label{ascond}
  \Phi(2-\nu,4;\rho) & \sim & \left \{
\begin{array}{ll}
  \frac{1}{\Gamma(2-\nu)} \rho^{-2-\nu} e^\rho, & \mbox{as}\; \rho
  \rightarrow \infty \\
1, & \mbox{as}\;\rho \rightarrow 0
\end{array}
\right. \\
\Psi(2-\nu,4;\rho) & \sim & \left \{
\begin{array}{ll}
  \rho^{\nu-2}, & \mbox{as}\; \rho \rightarrow \infty \\
  \frac{2}{\Gamma(2-\nu)} \rho^{-3}, & \mbox{as}\;\rho
  \rightarrow 0
\end{array}
\right. .
\end{eqnarray*}
We can adopt the same choice of $D_i$ also for $\nu=2,3,4,\ldots$,
being of no physical interest. Whence, one finally obtains:
\begin{eqnarray}
  \lefteqn{ \psi_n(r)= \frac{1}{2} \left(\frac{\nu}{\sqrt{nZ}}\right)^3
    \rho e^{-\rho/2} \left \{ \frac{1}{n!} \frac{d^n}{dt^n} \left.
    \left[ t^{\nu-2} (1-t)^{-\nu-2} \left(t- \frac{n+\nu}{2n}\right)^{n-1}
    \right] \right|_{t=t_0} \times \right.} \nonumber \hspace{75ex} \\
  \label{FINpsi}
  \times \int_0^{\frac{\nu-n}{2n}} e^{-\rho x} (-x)^{1-\nu}
  (1+x)^{1+\nu} dx - \nonumber \hspace{35ex} \\ - \left.
\begin{array}[t]{c}
  \mbox{res} \\ {\scriptstyle t=t_0}
\end{array}
\left [ e^{\rho t} t^{1-\nu} (1-t)^{1+\nu} \int_{t}^{\infty} t^{\nu-2}
  (1-t)^{-\nu-2} \frac{ \left(t- \frac{n+\nu}{2n}\right)^{n-1}} {
    \left(t- \frac{n-\nu}{2n}\right)^{n+1}} dt \right ] \right\}.
\end{eqnarray}
\begin{figure}[b]
  \centering \epsfig{file=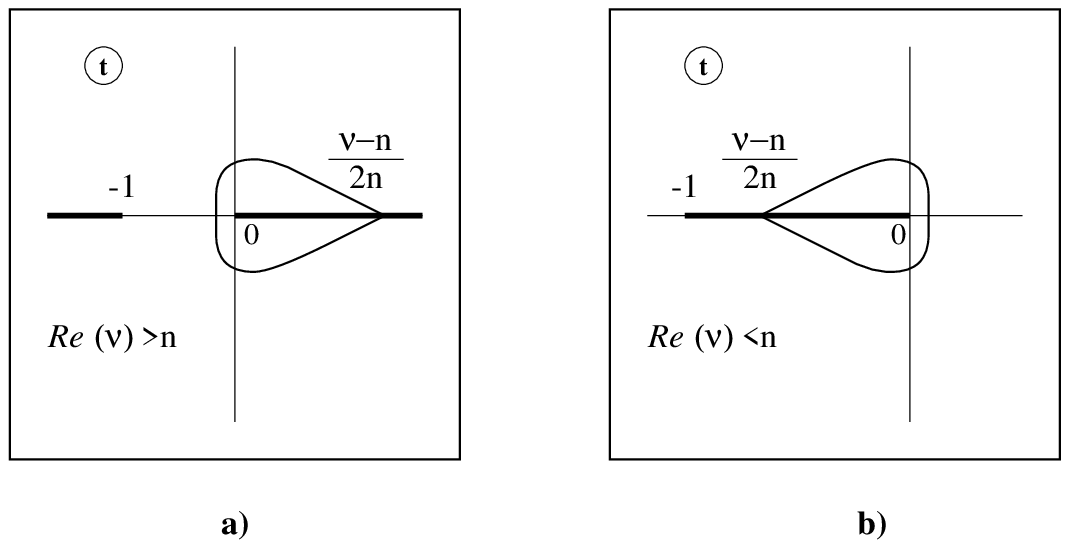}
\caption{The contour of integration in Eq.(32).}
\label{fig2}
\end{figure}
In the given derivation we assumed that parameter $\nu$ is subject to
condition: $n< \Re(\nu)<2,\;\Im \nu=0$. Analytical continuation of
(\ref{FINpsi}) on all $\nu$ is achieved by means of the substitution:
\begin{equation}
  \label{ancont1}
  \int_0^{\frac{\nu-n}{2n}} e^{-\rho x} (-x)^{1-\nu} (1+x)^{1+\nu} dx
  \rightarrow \left(1-e^{-2\pi \nu i}\right)^{-1}
  \int_{0+}^{\frac{\nu-n}{2n}} e^{-\rho x} (-x)^{1-\nu} (1+x)^{1+\nu}
  dx.
\end{equation}
Depending on the cases: $\Re \nu>n$ or $\Re \nu<n$, the integral in
the right-hand side here is taken along the paths shown in the
Figure \ref{fig2}(a),(b). Each of them starts at the point
$(-t_0)=(\nu-n)/(2n)$ lying on the lower edge of the corresponding cut,
encircles the origin in the clockwise (counterclockwise) sense, and
ends up at $(-t_0)$ lying on the upper edge of the same cut.
In the following we shall use for simplicity a sign of the ordinary
integral but imply, whenever necessary, the substitution
(\ref{ancont1}). Alternatively, the same analytical continuation can be
achieved by means of the identity:
\begin{eqnarray}
\lefteqn{
 \int_0^{\frac{\nu-n}{2n}} e^{-\rho x} (-x)^{1-\nu} (1+x)^{1+\nu} dx=}
\nonumber \hspace{60ex} \\
=\frac{(-1)^{1-\nu}}{2-\nu}\left(\frac{\nu-n}{2n}\right)^{2-\nu}
\Phi_1\left(2-\nu,-1-\nu,3-\nu,\frac{n-\nu}{2n},\frac{n-\nu}{2n}\rho\right),
\;\;\;\Re \nu <2.
  \label{DegHyper}
\end{eqnarray}
Here $\Phi_1(\ldots )$ denotes degenerate hypergeometric function of
two variables defined by the following series \cite{EH}:
\begin{equation}
  \label{degser}
  \Phi_1(\alpha,\beta,\gamma,x,y)= \sum_{m=0}^\infty \sum_{n=0}^\infty
  \frac{(\alpha)_{m+n} (\beta)_n}{(\gamma)_{m+n} m! n!} x^m
  y^n,\;\;\;|x|<1.
\end{equation}

In particular cases, $n=1,2,3$, explicit expressions for the functions
$\psi_n(r)$ are combined below.
\begin{eqnarray}
\label{psi1}
  \lefteqn{\underline{n=1}} \nonumber \hspace{40ex} \\[1ex]
\psi_1(r)=-\frac{32 \nu^4}{Z^{3/2}(\nu^2-1)^3 }
  \left(\frac{\nu-1}{\nu+1}\right)^\nu \rho e^{-\rho/2}
    \int_{0}^{\frac{\nu-1}{2}} e^{-\rho x} x^{1-\nu} (1+x)^{1+\nu} dx
    + \nonumber \\
+ \frac{2 \nu^3}{Z^{3/2}(\nu^2-1)}\rho e^{-\nu
      \rho/2} \\[3ex]
\label{psi2}
\lefteqn{\underline{n=2}} \nonumber \hspace{40ex} \\[1ex]
\psi_2(r)=-\frac{512 \sqrt{2} \nu^4}{Z^{3/2}(\nu^2-4)^3 }
\left(\frac{\nu-2}{\nu+2}\right)^\nu \rho
    e^{-\rho/2} \int_{0}^{\frac{\nu-2}{4}} e^{-\rho x} x^{1-\nu}
    (1+x)^{1+\nu} dx - \nonumber \\
-\frac{\sqrt{2} \nu^3}{2
      Z^{3/2}(\nu^2-4)^2}\rho e^{-\nu \rho/4} \left[\nu(\nu^2-4)\rho
    +4(\nu^2+4)\right]\hspace{1ex} \\[3ex]
\label{psi3}
\lefteqn{\underline{n=3}} \nonumber \hspace{40ex} \\[1ex]
\psi_3(r)=-\frac{864 \sqrt{3} \nu^4 \left(7\nu^2-27\right)}
{Z^{3/2}(\nu^2-9)^4 }
      \left(\frac{\nu-3}{\nu+3}\right)^\nu \rho e^{-\rho/2}
        \int_{0}^{\frac{\nu-3}{6}} e^{-\rho x} x^{1-\nu} (1+x)^{1+\nu}
        dx + \nonumber \\
+ \frac{\sqrt{3} \nu^3}{27 Z^{3/2}(\nu^2-9)^2}\rho e^{-\nu \rho/6}
        \left[\nu^2(\nu^2-9)\rho^2 -6\nu(\nu^2-27)\rho-(306\nu^2+486)\right]
\end{eqnarray}
It can be proved (see Appendix A) that Eqs.(\ref{psi1})-(\ref{psi3}) satisfy
orthogonality condition of Eq.(\ref{orthcond}).

Let us also give for reference explicit expression for the leading term in
the expansion of $\psi_n(r)$ in the small parameter $\nu/n \ll 1$:
\begin{eqnarray}
  \label{psiINF}
  \psi_n(r)
\begin{array}[t]{c}
  \asymp \\ {\scriptstyle n \gg \nu}
\end{array}
-2\left(\frac{\nu}{\sqrt{nZ}}\right)^3 \rho
\left[
16\nu \int_0^{1/2}e^{-\rho t}
\left(\frac{1}{2}-t\right)^{1-\nu}\left(\frac{1}{2}+t\right)^{1+\nu}dt +1
\right],\;\;\Re\nu<2.
\end{eqnarray}
This identity follows directly from Eq.(\ref{FINpsi}).

\subsection{Calculation of the light-shifts in particular cases}

Now we are in a position of calculating the matrix elements entering
Eq.(\ref{alpsi}) for $n=1,2$, being of particular interest. Integration, using
Eqs.(\ref{psi1}-\ref{psi2}), yields the following equivalent forms
of $\langle 1s||r||\psi_1\rangle,\;
\langle 2s||r||\psi_2\rangle$ (see Appendix B):
\begin{eqnarray}
\label{a11}
  \frac{1}{3} \langle 1s||r||\psi_1\rangle & = &
-\frac{512 \nu^9}{Z^4 (\nu^2-1)^2 (\nu+1)^8}
\int_0^1
\frac{t^{1-\nu}}{\left(1-\left(\frac{\nu-1}{\nu+1}\right)^2 t\right)^4} dt +
\frac{2\nu^2\left(2\nu^2-1\right)}{Z^4 \left(\nu^2-1\right)^2} \\
& \equiv & \frac{512 \nu^9}{Z^4 (\nu^2-1)^2 (\nu+1)^8 (\nu-2)}\,
\mbox{}_2F_1\left(4,2-\nu,3-\nu;\left(\frac{\nu-1}{\nu+1}\right)^2\right) +
\nonumber \\
\label{a12}
& & \hspace{45ex}
+ \frac{2\nu^2\left(2\nu^2-1\right)}{Z^4 \left(\nu^2-1\right)^2}, \\
\frac{1}{3} \langle 2s||r||\psi_2\rangle & = &
-\frac{2^{18}\nu^9}{Z^4 (\nu^2-4)^2(\nu+2)^8}
\int_0^1
\frac{t^{1-\nu}}
{\left(1-\left(\frac{\nu-2}{\nu+2}\right)^2 t\right)^4} dt +
\nonumber \\ \label{a21} & & \hspace{37ex}
+ 16\frac{\nu^2(\nu^4-64\nu^2+112)}
{Z^4 (\nu^2-4)^3} \\
& \equiv &
-\frac{2^{20}\nu^9}{Z^4 (\nu^2-4)(\nu+2)^{11}}
\int_0^1
\frac{t^{2-\nu}}
{\left(1-\left(\frac{\nu-2}{\nu+2}\right)^2 t\right)^5} dt +
\nonumber \\ \label{a22} & & \hspace{38ex}
+ 16\frac{\nu^2(5\nu^2+28\nu+28)}{Z^4 (\nu^2-4)(\nu+2)^2} \\
& \equiv &
\frac{2^{18} \nu^9}{Z^4 (\nu+2)^7 (\nu^2-4)^3}\,
\mbox{}_2F_1\left(4,2-\nu,3-\nu;\left(\frac{\nu-2}{\nu+2}\right)^2\right) +
\nonumber \\ \label{a23} & & \hspace{37ex}
+ 16 \frac{\nu^2(\nu^4-64\nu^2+112)}{Z^4 (\nu^2-4)^3} \\
& \equiv &
\frac{2^{18} \nu^9(\nu+1)}{Z^4 (\nu+2)^{12} (\nu-2)(\nu-3)}\,
\mbox{}_2F_1\left(4,3-\nu,4-\nu;\left(\frac{\nu-2}{\nu+2}\right)^2\right) +
\nonumber \\ \label{a24} & & \hspace{31ex}
+ 16 \frac{\nu^2(\nu^3+38\nu^2+84\nu+56)}{Z^4 (\nu-2)(\nu+2)^4}
\end{eqnarray}
Eqs.(\ref{a11}),(\ref{a21}) prove to be convenient for an analysis
of the DP's behavior when $\omega \gg Z^2/(2n^2)\equiv I_n$ and
$\omega \simeq 0$, as well as for calculation of the DP's imaginary part;
Eq.(\ref{a22}) is suitable for numerical calculation of the light-shift
of the $2s$-level; Eqs. (\ref{a12}),(\ref{a23}), and (\ref{a24}) define
the matrix elements in terms of the hypergeometric function \cite{AS},
$\mbox{}_2F_1(\ldots)$. In this form they easily admit analytical
continuation in $\nu$.

By means of Eqs.(\ref{alpsi}),(\ref{varch}) we can explicitly
express now DP, $\alpha_{1s}^S(\omega),\:\alpha_{2s}^S(\omega)$, as:
\begin{eqnarray}
  \label{DP1s}
\alpha_{1s}^S(\omega) & = & -\frac{1}{3}\left(
\left . \langle 1s||r||\psi_1\rangle
\right |_{\nu=\nu_{11}} +
\left . \langle 1s||r||\psi_1\rangle
\right |_{\nu=\nu_{12}} \right),\\
  \label{DP2s}
\alpha_{2s}^S(\omega) & = & -\frac{1}{3}\left(
\left . \langle 2s||r||\psi_2\rangle
\right |_{\nu=\nu_{21}} +
\left . \langle 2s||r||\psi_2\rangle
\right |_{\nu=\nu_{22}} \right).
\end{eqnarray}
Here we have introduced the following notations:
\begin{eqnarray}
  \label{nu1sPM}
 \nu_{11}=\frac{Z}{\sqrt{-2(\varepsilon_{1s}+\omega+i0)}},\;\;\;
  \nu_{12}=\frac{Z}{\sqrt{-2(\varepsilon_{1s}-\omega)}} \\
\label{nu2sPM}
 \nu_{21}=\frac{Z}{\sqrt{-2(\varepsilon_{2s}+\omega+i0)}},\;\;\;
  \nu_{22}=\frac{Z}{\sqrt{-2(\varepsilon_{2s}-\omega)}},
\end{eqnarray}
so that the following identities hold true:
\[
\omega^2=\frac{Z^4(1-\nu_{1m}^2)^2}{4\nu_{1m}^4} =
\frac{Z^4(4-\nu_{2m}^2)^2}{64\nu_{2m}^4},\;\;\;m=1,2.
\]
Here $\varepsilon_{1s}=-Z^2/2,\;\varepsilon_{2s}=-Z^2/8$ are the energies
of the $1s$- and $2s$-levels. We assume, according to the standard rule of
analytical continuation of the square root (see \cite{LL}), that
\renewcommand{\arraystretch}{1.5}
\begin{eqnarray}
  \label{anNU}
  \frac{Z}{\sqrt{-2(E+i0)}}=
\left\{
\begin{array}{ll}
\frac{Z}{\sqrt{-2E}}>0 & \mbox{if}\;E<0 \\
\frac{iZ}{\sqrt{2E}}   & \mbox{if}\;E>0
\end{array}
\right. .
\end{eqnarray}
\renewcommand{\arraystretch}{1}
Hence, by choosing the matrix elements
$\langle 1s||r||\psi_1\rangle,\;\langle 2s||r||\psi_2\rangle$ in the
forms of Eqs.(\ref{a12}),(\ref{a23}), DP can be expressed as:
\begin{eqnarray}
\lefteqn{
\alpha_{1s}^S(\omega) =- \frac{1}{\omega^2} - \sum_{m=1}^2
\frac{512 \nu_{1m}^9}{Z^4 (\nu_{1m}^2-1)^2 (\nu_{1m}+1)^8
(\nu_{1m}-2)}\, \times}
\nonumber \hspace{83.5ex} \\
  \label{DP1s1}
\times
\mbox{}_2F_1\left(4,2-\nu_{1m},3-\nu_{1m};
\left(\frac{\nu_{1m}-1}{\nu_{1m}+1}\right)^2
\right) , \\
\label{DP2s1}
\alpha_{2s}^S(\omega) = - \frac{1}{\omega^2} -\sum_{m=1}^2
\frac{2^{18} \nu_{2m}^9}{Z^4 (\nu_{2m}+2)^7(\nu_{2m}^2-4)^3}\,
\mbox{}_2F_1\left(4,2-\nu_{2m},3-\nu_{2m};
\left(\frac{\nu_{2m}-2}{\nu_{2m}+2}\right)^2
\right).
\end{eqnarray}
Eq.(\ref{DP1s1}) is in agreement with the well known result of Gavrila
\cite{Gav}, as it should.  Eqs.(\ref{DP1s1})-(\ref{DP2s1}) are rather
inconvenient, however, for numerical calculation of DP for energies
lying above the threshold of the levels, i.e. when parameters
$\nu_{11},\:\nu_{21}$ become purely imaginary (cf
Eq.(\ref{anNU}). Besides, they are also unsuitable for obtaining
various asymptotics of these quantities. As was already mentioned above,
for these purposes the integral forms of the matrix elements,
Eqs.(\ref{a11}),(\ref{a21}),(\ref{a22}), prove to be more
convenient. Below we combine various most important results
of such calculations (see Appendix for details).
\begin{itemize}
\item[1.] {\bf The case: $\boldmath \omega/Z^2 \ll 1$.}
\renewcommand{\arraystretch}{0.5}
\begin{eqnarray}
  \label{om1s0}
\alpha_{1s}^S(\omega)
\begin{array}[t]{c}
  \asymp \\ {\scriptstyle \omega/Z^2 \ll 1}
\end{array}
\frac{9}{2}\frac{1}{Z^4}+\frac{319}{12}\left(\frac{\omega}{Z^4}\right)^2
+ \ldots \hspace{3ex} \\
\label{om2s0}
\alpha_{2s}^S(\omega)
\begin{array}[t]{c}
  \asymp \\ {\scriptstyle \omega/Z^2 \ll 1}
\end{array}
120\frac{1}{Z^4}+21120\left(\frac{\omega}{Z^4}\right)^2 + \ldots
\end{eqnarray}
\item[2.] {\bf The case: $\boldmath \omega/Z^2 \gg 1$.}
\begin{eqnarray}
\lefteqn{
\alpha_{1s}^S(\omega)
\begin{array}[t]{c}
  \asymp \\ {\scriptstyle \omega/Z^2 \gg 1}
\end{array}
-\frac{1}{\omega^2}-\frac{4}{3}\frac{Z^4}{\omega^4} +
\frac{4\sqrt{2}}{3}(1+i)\frac{Z^5}{\omega^{9/2}} -
i\frac{4\pi}{3}  \frac{Z^6}{\omega^{5}} -}
\nonumber \hspace{77ex} \\
  \label{om1sINF}
-\frac{\sqrt{2}}{144}\left(
-336i-35i\pi^2+336+32\pi^2-3(1+i)\pi \ln(8Z^2/\omega)\right)
\frac{Z^7}{\omega^{11/2}} + \ldots \\
\lefteqn{
\alpha_{2s}^S(\omega)
\begin{array}[t]{c}
  \asymp \\ {\scriptstyle\omega/Z^2 \gg 1 }
\end{array}
-\frac{1}{\omega^2}-\frac{1}{6}\frac{Z^4}{\omega^4} +
\frac{\sqrt{2}}{6}(1+i)\frac{Z^5}{\omega^{9/2}} -
i\frac{\pi}{6} \frac{Z^6}{\omega^{5}} -}
\nonumber \hspace{77ex} \\
  \label{om2sINF}
-\frac{1}{2304}\sqrt{2}\left(
-504i-61i\pi^2+504+64\pi^2+3(1+i)\pi \ln(2Z^2/\omega)\right)
\frac{Z^7}{\omega^{11/2}} + \ldots
\end{eqnarray}
\item[3.] {\bf Calculation of $\boldmath \Im \alpha^S_{n}(\omega),\;
\omega>I_n \equiv Z^2/(2n^2)$.}
\begin{eqnarray}
\label{Im1s}
\Im \alpha^S_{1s}(\omega) & = &
\frac{64\pi}{3\omega^2}\:\frac{\eta^6}{(1+\eta^2)^3}
\frac{e^{-4 \eta \arctan(1/\eta)}}{1-e^{-2\pi\eta}},\;\eta=|\nu_{11}|,\\
\label{Im2s}
\Im \alpha^S_{2s}(\omega) & = &
\frac{2048\pi}{3\omega^2}\:\frac{\eta^6(1+\eta^2)}{(4+\eta^2)^4}
\frac{e^{-4 \eta \arctan(2/\eta)}}{1-e^{-2\pi\eta}},\;\eta=|\nu_{21}|.
\end{eqnarray}
Accordingly, the photoionization cross sections,
$\sigma^{(\gamma)}_{ns}(\omega) =
4\pi \alpha \omega \Im \alpha^S_{n}(\omega)$, are defined as:
\begin{eqnarray}
\label{sigma1s}
\sigma_{1s}^{(\gamma)}(\omega) & = &
\frac{2^9 \pi^2}{3Z^2}\alpha
\left(\frac{I_{1s}}{\omega}\right)^4
\frac{e^{-4 \eta \arctan(1/\eta)}}{1-e^{-2\pi\eta}},\;
\eta=\sqrt{\frac{I_{1s}}{\omega-I_{1s}}}>0,\\
\label{sigma2s}
\sigma_{2s}^{(\gamma)}(\omega) & = &
\frac{2^{14}\pi^2}{3Z^2}\alpha
\left(1+3\frac{I_{2s}}{\omega}\right)\left(\frac{I_{2s}}{\omega}\right)^4
\frac{e^{-4 \eta \arctan(2/\eta)}}{1-e^{-2\pi\eta}},\;
\eta=\sqrt{\frac{4I_{2s}}{\omega-I_{2s}}}>0.
\end{eqnarray}
\item[4.] {\bf The case: $\boldmath \omega/I_n \rightarrow 1+0$.}
\begin{eqnarray}
\lefteqn{
\frac{1}{3}\langle 1s||r||\psi_1\rangle
\begin{array}[t]{c}
  \asymp \\ {\scriptstyle \omega/I_{1s} \rightarrow 1+0}
\end{array}
\frac{88}{3Z^4}-\frac{256}{3Z^4}e^{-4}\mbox{Ei}(4)(1-i) - }
\nonumber \hspace{75ex} \\
  \label{om1sI}
-\frac{1}{9Z^4}
\left(1016-2816e^{-4}\mbox{Ei}(4)(1-i) \right) \frac{1}{\eta^2} + \ldots \\
\lefteqn{
\frac{1}{3}\langle 2s||r||\psi_2\rangle
\begin{array}[t]{c}
  \asymp \\ {\scriptstyle \omega/I_{2s} \rightarrow 1+0}
\end{array}
\frac{18896}{3Z^4}-\frac{131072}{3Z^4}e^{-8}\mbox{Ei}(8)(1-i) - }
\nonumber \hspace{75ex} \\
  \label{om2sI}
-\frac{1}{9Z^4}
\left(709952-4849664e^{-8}\mbox{Ei}(8)(1-i) \right) \frac{1}{\eta^2} + \ldots
\end{eqnarray}
Here parameters $\eta$ are defined by Eqs.(\ref{sigma1s}),(\ref{sigma2s});
$e=2.71828\ldots$; $\mbox{Ei}(\ldots)$ stands for the integral
exponential function; condition $\eta \gg 1$ is assumed in either case.
\item[5.] {\bf The case: $\boldmath Z^2/(2\omega n^2) \ll 1;\;
Z,\omega$ are fixed.}\\
By means of Eq.(\ref{psiINF}) and in view of the evident relations,
\begin{eqnarray*}
\nu_{m,2}
\begin{array}[t]{c}
  \asymp \\ {\scriptstyle n \gg 1}
\end{array}
\frac{Z}{\sqrt{2\omega}}\equiv \nu_0>0,\;\;\;
\nu_{m,1}
\begin{array}[t]{c}
  \asymp \\ {\scriptstyle n \gg 1}
\end{array}
\frac{iZ}{\sqrt{2\omega}}\equiv i\nu_0,
  \end{eqnarray*}
the leading term of the expansion of DP in
the small parameter, $\nu_0/n \ll 1$, can be expressed as
  \begin{eqnarray}
   \alpha_n^S(\omega)
\begin{array}[t]{c}
  \asymp \\ {\scriptstyle \nu_0/n \ll 1}
\end{array}
\frac{32 \nu^8}{Z^4} \frac{1}{n^3}
\left\{
 \Gamma(1-\nu_0) e^{-2\nu_0} \left[
4 \Psi(-\nu_0,1;4\nu_0) - \right. \right.
\nonumber \hspace{20ex} \\
    \label{nINF}
\left. \left. - (2\nu_0+1)\Psi(-\nu_0,0;4\nu_0)\right]
+ \left. (\ldots) \right|_{\nu_0 \rightarrow i\nu_0}
\right\}.
  \end{eqnarray}
Here $\Psi(\ldots)$ denotes irregular degenerate hypergeometric function
(cf (\ref{gensol})), whereas $\Gamma(\ldots)$ stands for the
$\Gamma$-function \cite{AS}.
\end{itemize}

\begin{table}[h]
\[
\renewcommand{\arraystretch}{1.5}
\begin{array}{||c|c||c|c||c|c||c|c||} \hline \hline
\multicolumn{4}{||c||}{1s-\mbox{\bf level}} &
\multicolumn{4}{|c||}{2s-\mbox{\bf level}}   \\ \hline \hline
\multicolumn{2}{||c||}{\omega_1= 3/16} &
\multicolumn{2}{|c||}{\omega_2= 1/8} &
\multicolumn{2}{|c||}{\omega_1= 3/16} &
\multicolumn{2}{|c||}{\omega_2= 1/8} \\ \hline \hline
\nu_{11} & \nu_{12} & \nu_{11} & \nu_{12} &
\nu_{21} & \nu_{22} & \nu_{21} & \nu_{22} \\ \hline
\sqrt{8/5} & \sqrt{8/11} & 2/\sqrt{3} & 2/\sqrt{5} &
i\sqrt{8} & \sqrt{8/5} & +i\infty & \sqrt{2} \\ \hline \hline
\end{array}
\]
\caption{The values of parameters $\nu_{ij},\;i,j=1,2$.}
\label{tabNU}
\end{table}
\begin{table}[h]
\[
\renewcommand{\arraystretch}{1.5}
\begin{array}{||c|c||c|c||} \hline \hline
\multicolumn{2}{||c||}{\alpha_{1s}^{S}(\omega)} &
\multicolumn{2}{|c||}{\alpha_{2s}^{S}(\omega)}   \\ \hline \hline
\omega_1 & \omega_2 & \omega_1 & \omega_2 \\ \hline \hline
-5.714105 & -4.962372 & 29.853542 - 12.823175i &
89.818540 - 46.045022i \\ \hline \hline
\end{array}
\]
\caption{The values of $\alpha_{ns}^{S}(\omega)$.}
\label{tabA}
\end{table}
Let us apply the results obtained to particular photon energies adopted
in the above-mentioned 1S-2S experiment in muonium atom ($Z=1$).
The latter is carried in the presence of two laser beams with the energies
$\omega_1=3/16\: \mbox{a.u.}\;(\lambda_1=244\:\mbox{nm}$) and
$\omega_1=1/8\: \mbox{a.u.}\;(\lambda_2=366\:\mbox{nm}$).
The corresponding values of $\nu_{ij},\;i,j=1,2$ (\ref{nu1sPM}),
(\ref{nu2sPM}) and $\alpha_{ns}^{S}(\omega)$ are compiled in the
Tables \ref{tabNU},\ref{tabA}. In obtaining $\alpha_{2s}^{S}(\omega_2)$
we used the value,
\[
\left. \frac{1}{3}\langle 2s||r||\psi_2\rangle
\right|_{\nu \rightarrow +i\infty} =
155.799140 - 46.045022i,
\]
being equal to the leading ($\eta$-independent) term in Eq.(\ref{om2sI}).
Its imaginary part coincides, as it should, with $\Im\alpha_{2s}^S(I_{2s})$
of Eq.(\ref{Im2s}), whereas the real part defines  the level shift at the
photoionization threshold. According to Eq.(\ref{sclshift}), the numbers
displayed enable to obtain, e.g. the following important dimensionless ratio:
\[
R_{\omega_1 \omega_2} \equiv
\frac{\Delta \varepsilon_{2s}(\omega_2)-\Delta \varepsilon_{1s}(\omega_2)}
     {\Delta \varepsilon_{2s}(\omega_1)-\Delta \varepsilon_{1s}(\omega_1)}
\equiv \frac{I_{\omega_2}}{I_{\omega_1}}
\frac{\Re\alpha^S_{2s}(\omega_2)-\Re\alpha^S_{1s}(\omega_2)}
 {\Re\alpha^S_{2s}(\omega_1)-\Re\alpha^S_{1s}(\omega_1)}=
2.664 \frac{I_{\omega_2}}{I_{\omega_1}}.
\]
Here $I_{\omega_1}\equiv P_{\omega_1}/S_1,\;
I_{\omega_2}\equiv P_{\omega_2}/S_2$
are the beam intensities, $P_{\omega_1,\omega_2}$, $S_{1,2}$
being the corresponding powers and cross sections. The absolute value of
the light-shift of the $ns$-level due to (one) photon beam of the field
strength, $E_{\omega}$, is defined as (henceforth in this section the
ordinary units are used)
\[
\Delta \varepsilon_{ns}(\omega)=
\frac{1}{4}E_{\omega}^2 \Re \alpha_{ns}^S(\omega),\;\;\;
E_{\omega}\equiv\sqrt{\frac{2 I_{\omega}}{c \varepsilon_0}}=
5.338 \cdot 10^{-5} \sqrt{\frac{A_p^\omega}{S_\omega \tau_\omega}} E_0 \;
\mbox{mm}\cdot \mbox{ns}^{-1/2} \cdot \mbox{mJ}^{-1/2}.
\]
Here $A_p^\omega$ and $\tau_\omega$ stand for the energy of the beam
within one pulse (in {\em mJ}) and the pulse duration (in {\em ns});
the beam cross section, $S_\omega$,  is measured in $mm^2$;
$E_0=m_e^2 e^5/\hbar^4=5.142 \cdot 10^{11}\:\mbox{Volts/m}$ denotes the
atomic unit of electric field strength.
For the typical values of these parameters adopted in the experiment
($S_\omega=2\times 3 \:\mbox{mm}^2,\;\tau_\omega=28\: \mbox{ns},\;
A_p^\omega=6\:\mbox{mJ}$)
one obtains: $E_{\omega}\simeq 1.0 \cdot 10^{-5} E_0$, i.e. the electric
field employed happens to be rather weak. Accordingly, the {\em average\/}
intensity within a pulse equals:
to $I_\omega = A_p^\omega/(S_\omega \tau_\omega) \simeq 3.57 \cdot 10^6\:
\mbox{W}/\mbox{cm}^2$. In the presence of two counterpropagating beams
\footnote{The presence of {\em two\/} counterpropagating beams
of the same frequency enables to avoid both the Doppler-broadening
and the Doppler-shift of the line \cite{CTandD1972}.} having  the intensities,
$I_{\omega},\:I'_{\omega}$, and the same frequency, $\omega$, the shift
of the level (at the frequency $\omega$) can be expressed in the form:
\begin{eqnarray*}
\Delta \varepsilon_{ns}(\omega) & = &
\frac{1}{4}\left(E_{\omega}^2 + E_{\omega}^{'2}\right)
\Re \alpha_{ns}^S(\omega) \equiv
4.6875 (I_{\omega}+I'_{\omega}) \Re \alpha_{ns}^S(\omega) \\
& \equiv & 4.6875
\left(\frac{A_p^\omega}{S_\omega \tau_\omega} +
\frac{A_p^{'\omega}}{S'_\omega \tau'_{\omega}} \right)
\Re \alpha_{ns}^S(\omega)\:a_0^{-3} \;\mbox{mm}^2 \cdot \mbox{ns}^{-1}
\cdot \mbox{mJ}^{-1} \cdot \mbox{MHz}.
\end{eqnarray*}
Here it is taken into account that $\alpha_{ns}^S(\omega)$, whose values
are dislayed in the Table (\ref{tabA}), is measured in the units of $a_0^3$,
$a_0=\hbar^2/(m_e e^2)=0.529 \cdot 10^{-8}\:\mbox{cm}$
being the Bohr radius. As a result, the {\em total\/} energy shift between
$ns$- and $ms$-levels caused by two counterpropagating beams of the same
frequency $\omega$ is, then, given by
\begin{eqnarray}
\label{SHIFTnm1}
\Delta {\cal E}_{nm}(\omega) & \equiv &
\Delta \varepsilon_{ns}(\omega) - \Delta \varepsilon_{ms}(\omega) \\
& = & 4.6875  \left(
\frac{A_p^\omega}{S_\omega \tau_\omega} + \frac{A_p^{'\omega}}
{S'_\omega \tau'_\omega} \right)
\left(\Re \alpha_{ns}^S(\omega)-\Re \alpha_{ms}^S(\omega)\right)a_0^{-3}
\;\mbox{mm}^2 \cdot \mbox{ns}^{-1} \cdot \mbox{mJ}^{-1} \cdot \mbox{MHz}.
\nonumber
\end{eqnarray}
An application of this formula to the case of 1S-2S experiment, assuming
that parameters of the counterpropagating beams are identical
($S_{\omega_1}=S'_{\omega_1}=S_{\omega_2}=S'_{\omega_2}=
2\times 3 \:\mbox{mm}^2,\\
\tau_{\omega_1}=\tau'_{\omega_1}=\tau_{\omega_2}=\tau'_{\omega_2}=
28\: \mbox{ns},\;
A^{\omega_1}_p=A^{'\omega_1}_p=
A^{\omega_2}_p=A^{'\omega_2}_p=
6\:\mbox{mJ}$), yields:
\begin{eqnarray}
\label{num1s}
\Delta {\cal E}_{21}(\omega_1) & = & 11.9\:\mbox{MHz} \\
\label{num2s}
\Delta {\cal E}_{21}(\omega_2) & = &
R_{\omega_1 \omega_2} \Delta {\cal E}_{21}(\omega_1)=31.7\:\mbox{MHz}.
\end{eqnarray}

\section{Conclusion}

The numbers of Eqs.(\ref{num1s}),(\ref{num2s}) may serve as a good
illustration of the method employed. They are, however, of independent
significance. The value of $\Delta {\cal E}_{21}(\omega_1)$ is in fair
agreement with the result \cite{Beausoleil}. One has to stress that in
obtaining these energy shifts we used the {\em average}
intensities. As it shown in \cite{Beausoleil}, an account for a
space-inhomogeneity of a laser field may increase each of these
numbers by a factor of $10$.  It is relevant to point out here that
the results obtained for $n=1,2$ can be extended on the case of
arbitrary $n$.  Such type of calculation can be mostly efficiently
performed on the basis of Eq.(\ref{FINpsi}) valid for all $n$,
e.g. with the help of Maple \cite{Maple}, being an easy-to-use
computer algebra program. Hence, in the contrast to a pure numerical
scheme (say, in a fashion of Beausoleil \cite{Beausoleil}), one
obtains analytical, rather than numerical, result which is already
well adapted for further numerical calculation (one-fold
integration). Such an integration is performed only at the
very last stage of calculation and usually carried out in no time.
Furthermore, by that means one considerably reduces numerical
errors. Beyond that, analytical formulae admit straightforward
computation of the various asimptotics with respect to all parameters
encounted in them. As an example we can mention
the result of Eqs.(\ref{om1sI})-(\ref{om2sI}) when the photon energy,
$\omega$, tends to the threshold, $I_{ns}$, being of particular importance
for the problem under consideration.  This
argument may be considered as additional advantage of the method
employed. Besides, as was already mentioned in the Introduction, the
formula (\ref{FINpsi}), as well as its particular cases,
Eqs.(\ref{psi1})-(\ref{psi3}), enable a straightfoward calculation of
the sums, Eq.(\ref{gensum}). In fact, $S_n^{(\mu)}(\omega)$ can be
defined, in analogy with (\ref{alpsi}), as
\[
S_n^{(\mu)}(\omega)=
\langle ns||r^\mu||\psi_n(\varepsilon_{ns}+\omega+i0) \rangle +
\langle ns||r^\mu||\psi_n(\varepsilon_{ns}-\omega) \rangle.
\]
The analysis, whose details will be given elsewhere, shows that for
low $n$ it can be expressed in a closed form for any $\mu$. The
corresponding calculations prove to be simpler than those where
the Green function is employed. We would interpret this circumstance,
thereby, as a self-contained importance of the current method when it
is applied to the problem under consideration. One would expect that
this advantage will even be enhanced if the states with $n \geq 4$ are
taken into consideration. It is interesting to emphasize here that one
we can easily estimate the speed of the DP's decrease, as $n
\rightarrow \infty$. Namely, according to Eq.(\ref{nINF}),
$\alpha_n^S(\omega) \asymp C/n^3,\:n/\nu_0 \gg 1$. So that the
threshold itself ($E=0$), being the limit: $\varepsilon_{ns} \rightarrow 0,\;
\mbox{as}\:n \rightarrow \infty$, is not affected by the laser field.
This is in agreement with the well known result due to Ritus \cite{Ritus}
stating that the DP vanishes for continuum states. The $ns-$levels
with $n \geq 4$ prove to be of particular importance owing to
extensive experimental, as well as theoretical, investigation of the
spectroscopic properties of the few-body systems, being carried out at
present. In particuar, we consider the light-shift calculation in the
$3s-,\:4s-,\:5s-$levels as a subject of future publications.

\newpage
\section{Acknowledgment}

One of us (V.Y.), being an Alexander von Humboldt Research Fellow, wishes
to acknowledge his gratitude to the Alexander von Humboldt Foundation for
financial support. Besides, it would be a pleasure for him to thank
Dr. A. Korol, Prof. V.K. Ivanov, Prof. S. Sheinerman, and Dr. L. Gerchikov,
as well as all paricipants of a theoretical seminar at the A.F. Ioffe
Physico-Technical Insitute (St. Petersburg, Russia), for stimulation
discussion on the problem. This work has been also funded in part by the
Grants NWI300-NWI300 from the International Science Foundation.
\newpage
\section{Appendix}
\renewcommand{\thesubsection}{\Alph{subsection}}
\renewcommand{\theequation}{\Alph{subsection}.\arabic{equation}}
\subsection{Orthogonality-relations}
\setcounter{equation}{0}

Let us prove, say for $n=1,2$, that the functions (\ref{psi1})-(\ref{psi3})
satisfy ``orthogonality relation'', Eq.(\ref{orthcond}). Namely,
on calculating the integral,
\[
\langle \psi_n|2p\rangle \equiv
\frac{Z^{3/2}}{2\sqrt{6}} \int_0^\infty r^3 e^{-Zr/2} \psi_n(r) dr,\;
n=1,2,
\]
we obtain:
\begin{eqnarray}
  \label{qtest1s1}
\langle \psi_1|2p\rangle =
-\frac{4\sqrt{6}\nu^8}{Z^3 \left(\nu^2-1\right)^3}
\left(\frac{\nu-1}{\nu+1}\right)^\nu
\int_0^{\frac{\nu-1}{2}}\frac{x^{1-\nu}(1+x)^{1+\nu}}
{\left(x+\frac{\nu+2}{4}\right)^5} dx +
\frac{256}{243} \frac{\sqrt{6}\nu^2}{Z^3 \left(\nu^2-1\right)},
\hspace{3ex} \\
  \label{qtest2s1}
\langle \psi_2|2p\rangle =
-\frac{128\sqrt{3}\nu^8}{Z^3 \left(\nu^2-4\right)^3}
\left(\frac{\nu-2}{\nu+2}\right)^\nu
\int_0^{\frac{\nu-2}{4}}\frac{x^{1-\nu}(1+x)^{1+\nu}}
{\left(x+\frac{\nu+2}{4}\right)^5} dx -
\frac{8 \sqrt{3}\nu^2 \left(7\nu^2-12\right)}{Z^3 \left(\nu^2-4\right)^2}.
\end{eqnarray}
After the variable changes, $x/(x+1)=t (\nu-1)/(\nu+1),\;
x/(x+1)=t (\nu-2)/(\nu+2)$, Eqs.(\ref{qtest1s1}),(\ref{qtest2s1})
take the final form:
\begin{eqnarray}
  \label{qtest1s2}
\langle \psi_1|2p\rangle & = &
-\frac{2^{12}\sqrt{6}\nu^8}{Z^3 (\nu-1)(\nu+1)^5(\nu+2)^5}
\int_0^1
\frac{t^{1-\nu} \left(1-\frac{\nu-1}{\nu+1}t\right)}
{\left(1-\frac{(\nu-1)(\nu-2)}{(\nu+1)(\nu+2)}t\right)^5} dt +
\frac{256}{243} \frac{\sqrt{6}\nu^2}{Z^3 \left(\nu^2-1\right)}
\nonumber \\
& = & \frac{1024}{243}\frac{\sqrt{6}\nu^2}{Z^3 \left(\nu^2-4\right)}
\equiv \frac{\langle 2p||r||1s\rangle}{E-\varepsilon_{2p}}, \\
  \label{qtest2s2}
\langle \psi_2|2p\rangle & = &
-\frac{2^{17}\sqrt{3}\nu^8}{Z^3 (\nu-2)(\nu+2)^{10}}
\int_0^1
\frac{t^{1-\nu} \left(1-\frac{\nu-2}{\nu+2}t\right)}
{\left(1-\left(\frac{\nu-2}{\nu+2}\right)^2 t\right)^5} dt -
8 \frac{\sqrt{3}\nu^2 \left(7\nu^2-12\right)}{Z^3 \left(\nu^2-4\right)^2}
\nonumber \\
& = & -24\frac{\sqrt{3}\nu^2}{Z^3 \left(\nu^2-4\right)}
\equiv \frac{\langle 2p||r||2s\rangle}{E-\varepsilon_{2p}}.
\end{eqnarray}
Here we have used the integrals:
\begin{eqnarray}
  \label{i1s}
\int_0^1
\frac{t^{1-\nu} \left(1-\frac{\nu-1}{\nu+1}t\right)}
{\left(1-\frac{(\nu-1)(\nu-2)}{(\nu+1)(\nu+2)}t\right)^5} dt & = &
-\frac{1}{1296}\frac{(1+\nu)^4(\nu+2)^4}{\nu^4(\nu-2)},\\
\int_0^1
\frac{t^{1-\nu} \left(1-\frac{\nu-2}{\nu+2}t\right)}
{\left(1-\left(\frac{\nu-2}{\nu+2}\right)^2 t\right)^5} dt & = &
-\frac{1}{4096}\frac{(2+\nu)^8}{\nu^4(\nu-2)}.
\end{eqnarray}
These can be calculated by means of the following elementary relation:
\begin{equation}
  \label{recrel1}
  \int_0^1\frac{t^{1-\nu}(1-at)}{(1-bt)^{5}}dt=
\frac{1}{4}\frac{b-a}{b(1-b)^4} +\frac{4a+(2+\nu)(b-a)}{4b}
\int_0^1\frac{t^{1-\nu}}{(1-bt)^4}dt.
\end{equation}
Eqs. (\ref{qtest1s2}),(\ref{qtest2s2}) imply that
$\psi_1(r),\;\psi_2(r)$ satisfy Eqs.(\ref{orthcond}), as they should.
The case of arbitrary $n$ can be treated accordingly.

\subsection{Matrix elements of $\psi_n(r)$}
\setcounter{equation}{0}

Let us give here some details of a derivation of Eqs.(\ref{a11})-(\ref{a23}).
On calculating the integrals, $\langle ns||r||\psi_1\rangle,\;n=1,2$, using
Eqs.(\ref{psi1}),(\ref{psi2}) and explicit expressions \cite{LL},
\begin{equation}
  \label{R1s2s}
R_{1s}(r)=2Z^{3/2}e^{-Zr},\;\;\;
R_{2s}(r)=\frac{Z^{3/2}}{\sqrt{2}} \left(1-\frac{Zr}{2}\right)e^{-Zr/2},
\end{equation}
we get:
\begin{eqnarray}
\label{qin1s1}
  \frac{1}{3} \langle 1s||r||\psi_1\rangle =
-32\frac{\nu^8}{Z^4 (\nu^2-1)^3}\left(\frac{\nu-1}{\nu+1}\right)^\nu
\int_0^{\frac{\nu-1}{2}}
\frac{x^{1-\nu} (1+x)^{1+\nu}}
{\left(x+\frac{\nu+1}{2}\right)^5} dx +
\frac{2\nu^2}{Z^4 \left(\nu^2-1\right)}, \hspace{2ex} \\
\label{qin2s1}
\lefteqn{
  \frac{1}{3} \langle 2s||r||\psi_2\rangle =
-256\frac{\nu^8}{Z^4 (\nu^2-4)^3}\left(\frac{\nu-2}{\nu+2}\right)^\nu
\int_0^{\frac{\nu-2}{4}}
\frac{x^{1-\nu} (1+x)^{1+\nu}}
{\left(x+\frac{\nu+2}{4}\right)^5} dx +} \hspace{80ex} \nonumber \\
+320\frac{\nu^9}{Z^4 (\nu^2-4)^3}\left(\frac{\nu-2}{\nu+2}\right)^\nu
\int_0^{\frac{\nu-2}{4}}
\frac{x^{1-\nu} (1+x)^{1+\nu}}
{\left(x+\frac{\nu+2}{4}\right)^6} dx -
16\frac{\nu^2(28-13\nu^2)}{Z^4 (\nu^2-4)^2}.
\end{eqnarray}
After making the variable changes, $x/(x+1)=t (\nu-1)/(\nu+1),\;
x/(x+1)=t (\nu-2)/(\nu+2)$, Eqs.(\ref{qin1s1}),(\ref{qin2s1})
take the form:
\begin{eqnarray}
\label{qin1s2}
\frac{1}{3} \langle 1s||r||\psi_1\rangle =
-\frac{2^{10}\nu^8}{Z^4 (\nu^2-1)(\nu+1)^9}
\int_0^1
\frac{t^{1-\nu} \left(1-\frac{\nu-1}{\nu+1}t\right)}
{\left(1-\left(\frac{\nu-1}{\nu+1}\right)^2 t\right)^5} dt +
\frac{2\nu^2}{Z^4 \left(\nu^2-1\right)}, \hspace{8ex} \\
\lefteqn{
\frac{1}{3} \langle 2s||r||\psi_2\rangle =
-\frac{2^{18}\nu^8}{Z^4 (\nu^2-4)(\nu+2)^9}
\int_0^1
\frac{t^{1-\nu} \left(1-\frac{\nu-2}{\nu+2}t\right)}
{\left(1-\left(\frac{\nu-2}{\nu+2}\right)^2 t\right)^5} dt +}
\hspace{80ex} \nonumber \\
\label{qin2s2}
+\frac{2^{17}\nu^9}{Z^4 (\nu^2-4)(\nu+2)^{10}}
\int_0^1
\frac{t^{1-\nu} \left(1-\frac{\nu-2}{\nu+2}t\right)^2}
{\left(1-\left(\frac{\nu-2}{\nu+2}\right)^2 t\right)^6} dt -
16\frac{\nu^2(28-13\nu^2)}{Z^4 (\nu^2-4)}.
\end{eqnarray}
Finally, on applying successfully recurrence relations,
\begin{eqnarray}
\lefteqn{
  \int_0^1\frac{t^{1-\nu}(1-at)^2}{(1-bt)^6}dt=
\frac{(b-a)^2}{b^2}\int_0^1\frac{t^{1-\nu}}{(1-bt)^6}dt +
2\frac{a(b-a)}{b^2}\int_0^1\frac{t^{1-\nu}}{(1-bt)^5}dt +}
\nonumber \hspace{78ex}\\
\label{recrel2}
+\frac{a^2}{b^2}\int_0^1\frac{t^{1-\nu}}{(1-bt)^4}dt, \\
\label{recrel3}
  \int_0^1\frac{t^{1-\nu}}{(1-bt)^{n+1}}dt=
\frac{1}{n(1-b)^n} -\frac{2-\nu-n}{n}
\int_0^1\frac{t^{1-\nu}}{(1-bt)^n}dt, \hspace{19ex}
\end{eqnarray}
together with (\ref{recrel1}) to Eqs.(\ref{qin1s2}),(\ref{qin2s2}),
we retrieve identities (\ref{a11}),(\ref{a21}).

\end{document}